\documentclass[a4paper, amsfonts, amssymb, amsmath, reprint, showkeys, 
nofootinbib, twoside, twocolumn, citeautoscript]{revtex4-2}
\usepackage[colorinlistoftodos, color=green!40, prependcaption]{todonotes}
\usepackage[colorlinks=true, urlcolor=blue, linkcolor=red]{hyperref}
\usepackage{array}
\usepackage[scr=rsfso]{mathalfa} 
\usepackage{amsmath}

\begin{document}
\title{Exploring Fourier methods with beer bottles}
\author{David Kordahl}
    \email[Correspondence email address: ]{dkordahl@centenary.edu}
    \affiliation{Centenary College of Louisiana, Shreveport, LA, USA}
\author{Emma Foster}
	\affiliation{Centenary College of Louisiana, Shreveport, LA, USA}

\date{} 

\begin{abstract}
As anyone who has blown across the mouth of a beer bottle knows, beer bottles have a well-defined fundamental frequency. This paper shows how a beer bottle’s acoustical resonance can be modeled as a one-dimensional driven-damped oscillator and includes enough detail to be useful in undergraduate laboratory experiments. While the frequency-domain Green's function of the bottle can be extracted through sequential pure-tone measurements, sufficient data to fit the model's parameters can be collected in just a few seconds when Fourier methods are used.
\end{abstract}

\keywords{Acoustics, Driven Damped Oscillator, Fourier Methods}

\maketitle

\section{Introduction}

In 1979, the \textit{American Journal of Physics} (\textit{AJP}) published ``The great beer bottle experiment," which described a method for allowing introductory physics students to measure the speed of sound using tuning forks and a beer bottle \cite{Smith1979}. The authors noted that the experiment was successful, but added a caveat. ``The experiment is well received by our students for the wrong reasons (imagine entering a lab, finding that your apparatus consists of dozens of beer bottles, and being told that the lab demonstrators emptied them all last night)." This article also describes acoustical experiments involving beer bottles, which we also hope will be well received by students---even if for equally wrong reasons. 

The experiments presented below follow a suggestion from another \textit{AJP} article. In 2016, Wilkinson \textit{et al.} \cite{Wilkinson2016} showed how soda cans may be modeled as driven-damped oscillators. They used pure tones to measure a can's acoustical response using steady-state amplitudes, but mentioned that the cavity's acoustical response might also be obtained from shorter bursts of sound analyzed via a fast Fourier transform (FFT). In the present article, we apply that suggestion to beer bottles. We first review the correct treatment of the steady-state case, and then demonstrate how the bottle's acoustical response may be inferred using FFTs.

The experiments described here are simple enough that they may be readily performed by undergraduates. They use fundamental concepts from the undergraduate curriculum, including driven oscillations, Green's functions---and, of course, Fourier transforms. 

While Fourier methods are ubiquitous in physics, there is no single standard way to introduce them to undergraduate students. Some first encounter them in mathematical methods coursework \cite{Boas2006}, while others might first encounter them during a course in electromagnetism \cite{Griffiths2017}. The experiments here are designed to complement the discussions of periodic forcing in an analytical mechanics course \cite{Taylor2005}. They investigate a situation that is not quite obvious, but for which it is easy to carry out both experiments and calculations. 

\begin{figure}
\centering
\includegraphics[width=0.5\textwidth]{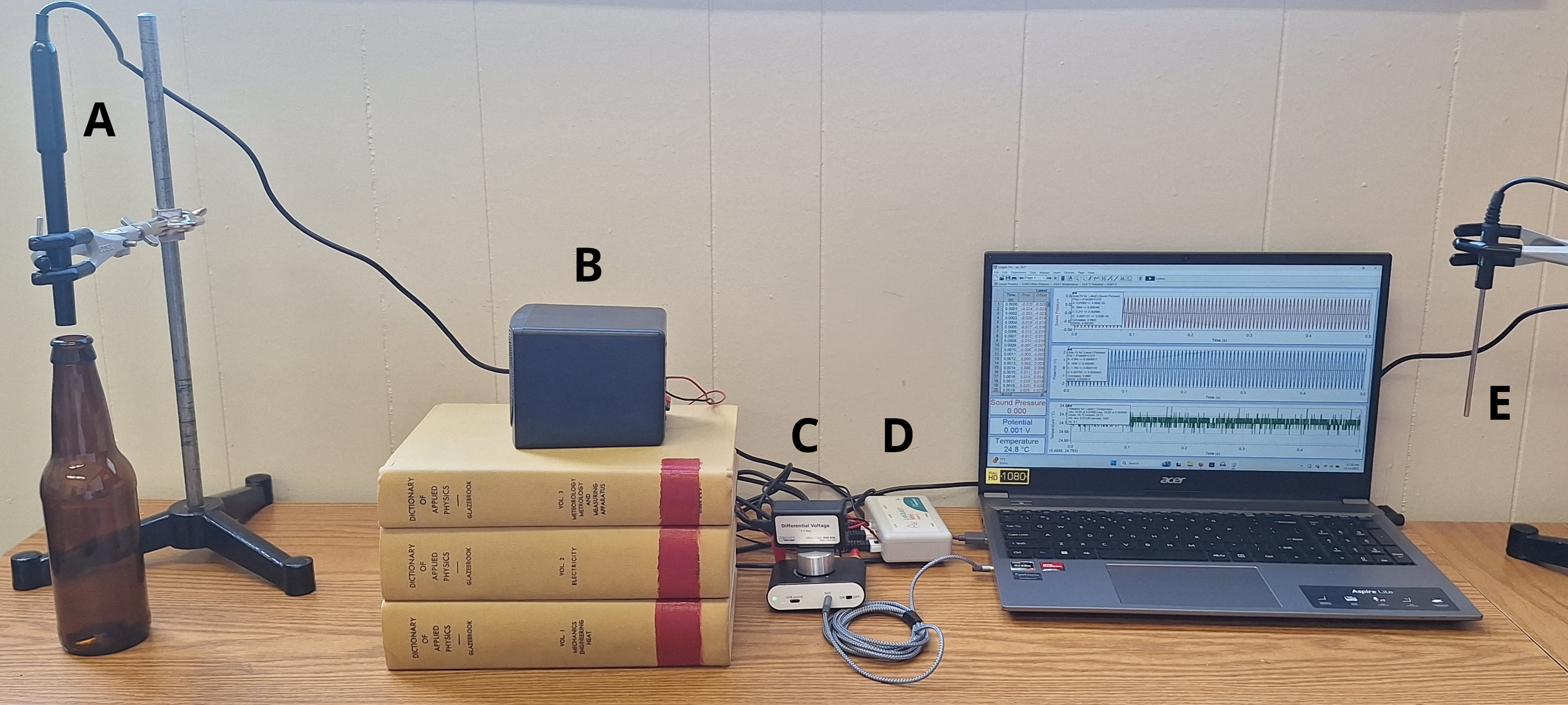}
\caption{Experimental setup. A microphone (A) measures the signal from a speaker (B), which is driven by an amplifier (C). (The stack of books boosts the speaker up to the microphone height.) Time-series data from the amplifier voltage and the microphone signal are measured and fed into the computer (D), which may also monitor the room temperature via an optional thermometer (E). Each measurement is taken both with the bottle below the microphone, and without the bottle.}\label{fig:bottle/no bottle setup}
\end{figure}

Our setup is straightforward and inexpensive, as shown in Fig.~\ref{fig:bottle/no bottle setup}. Signals are generated using the open-source audio software, Audacity \cite{Audacity}, and ported via a headphone cable into a speaker amplifier (\$34.00), which powers a passive desktop speaker (\$35.00). The amplified signal is measured by a Vernier Differential Voltage Probe (\$49.00), and the sound is measured by a Vernier Microphone (\$55.00), which connects to the computer using a Vernier LabQuest Mini (Model 2 costs \$189.00). Vernier devices make data collection easy, but one could always use other hardware. In the measurements below, a single 12 oz. heritage beer bottle has been used.

After reviewing in Sec.~\ref{sec:theory} how the driven-damped oscillator model applies to this setup, this paper gives three different approaches to recovering the frequency-domain Green's function $G(\omega)$ experimentally. Sec.~\ref{sec:pure tones} shows how the phase and amplitude of $G(\omega)$ can be extracted using pure sinusoidal tones, improving upon the approach of Wilkinson \textit{et al.} Sec.~\ref{sec: chirp tones} then shows how the same information can be obtained with just a few seconds' worth of data using Fourier methods and chirp signals. The subsections of Sec.~\ref{sec: chirp tones} describe two different possible exercises, one (\ref{subsec:incoherent}) which extracts resonance parameters using only FFT magnitudes, and the other (\ref{subsec:coherent}) which also employs the FFT phase. Sec.~\ref{sec:conclusion} concludes with suggested extensions to this work.

\section{Basic Theory}\label{sec:theory}

The standard physical model of acoustical resonance advanced by Helmholtz \cite{Helmholtz} considers a volume of air contained in the mouth of a bottle that is pushed back and forth by oscillating pressure differences between the inside and the outside of the bottle. The plug's resonance frequency $\omega_0$ can be estimated in terms of the bottle's volume and the opening's cross-sectional area \cite{HelmholtzPhysics}. When this oscillation is treated as driven and damped, two more parameters are introduced \cite{Wilkinson2016}. The first reflects that the bottle's pressure oscillations are driven and are coupled to the outside pressure oscillations via a dimensionless parameter $\alpha$. The second captures the fact that the bottle's pressure oscillations are damped, and, in the absence of outside forces, will decay in amplitude as $e^{-\beta t}$. 

The dynamical equation for $p_B(t)$, the pressure contribution of the bottle at the location of the microphone, should include a restoring force proportional to $\omega_0^2$, an external force proportional to $p_S(t)$, the pressure contribution of the nearby speaker, and a damping force proportional to $\beta$. It is a lightly disguised version of Newton's second law:
\begin{equation}\label{eq:damped SHO equation of motion}
\ddot{p}_B(t) = -\underbrace{\omega_0^2 p_B(t)}_{\substack{\text{restoring}\\ \text{force}}} + \underbrace{2 \alpha \beta \omega_0 p_S(t)}_{\substack{\text{driving}\\ \text{force}}} - \underbrace{2 \beta \dot{p}_B(t)}_{\substack{\text{damping}\\ \text{force}}} .
\end{equation}

To solve this, we can use the Fourier transform
\begin{equation}
\mathscr{F}\left[ f(t) \right] = \tilde{f}(\omega) = \int_{-\infty}^{+\infty} f(t) e^{-i\omega t} dt,
\end{equation}
and its corresponding inverse transform
\begin{equation}
\mathscr{F}^{-1}\left[ \tilde{f}(\omega) \right] = f(t) = \frac{1}{2 \pi} \int_{-\infty}^{+\infty} \tilde{f}(\omega) e^{+i\omega t} d\omega.
\end{equation}
If we take the Fourier transform of Eq.~\ref{eq:damped SHO equation of motion}, we find
\begin{equation}\label{eq:FT of damped SHO equation of motion}
-\omega^2 \tilde{p}_B(\omega)  = - 2 i \omega \beta \tilde{p}_B(\omega) + 2 \alpha \beta \omega_0 \tilde{p}_S(\omega) - \omega_0^2 \tilde{p}_B(\omega).
\end{equation}
From this, we can solve for $\tilde{p}_B(\omega)$ as
\begin{equation}\label{eq:solution for x(omega)}
\tilde{p}_B(\omega) = \left(\frac{2 \alpha \beta \omega_0 }{(\omega_0^2-\omega^2) + 2 i \beta \omega} \right) \tilde{p}_S(\omega).
\end{equation}

\begin{figure}
\centering
\includegraphics[width=0.5\textwidth]{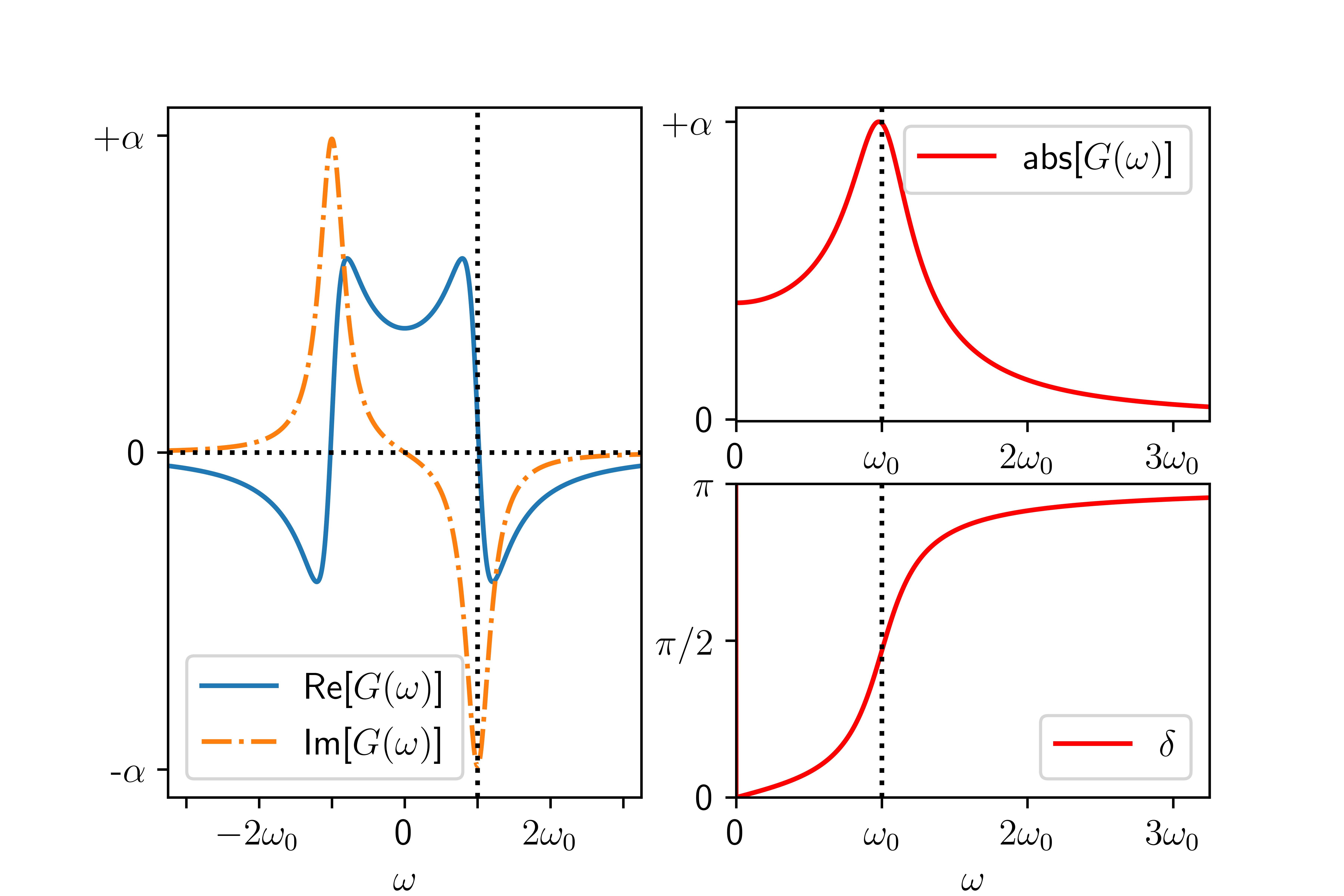}
\caption{\textit{Left}: Real and imaginary parts of $G(\omega)$. \textit{Right top}: Magnitude of $G(\omega)$. \textit{Right bottom}: Phase of $G(\omega)$. An unrealistically large $\beta = \omega_0/5$ has been used for plotting; typically, $\beta \ll \omega_0$, making the peaks in $G(\omega)$ much narrower.}\label{fig:green's function for oscillator - theory}
\end{figure}

This has the form of a Green's function \cite{Challis2003} 
\begin{equation}\label{eq:Green's function generic}
\tilde{p}_B(\omega) = G(\omega) \tilde{p}_S(\omega)
\end{equation}
with 
\begin{equation}\label{eq:Green's function specific}
G(\omega) = \frac{2 \alpha \beta \omega_0}{(\omega_0^2-\omega^2) + 2 i \beta \omega}.
\end{equation}
$G(\omega)$ can be plotted via its real and imaginary parts, or it can be represented \cite{Taylor2005} in terms of an amplitude
\begin{equation}\label{eq:G(omega) amplitude prediction}
|G(\omega)| = \frac{2 \alpha \beta \omega_0}{\sqrt{(\omega_0^2-\omega^2)^2+4\beta^2 \omega^2}}
\end{equation}
and phase
\begin{equation}\label{eq:G(omega) phase shift prediction}
\delta(\omega) = \arctan\left(\frac{2 \beta \omega}{\omega_0^2 - \omega^2}\right).
\end{equation}
such that
\begin{equation}
G(\omega) = |G(\omega)| e^{-i \delta(\omega)}.
\end{equation}
Both forms are shown in Fig.~\ref{fig:green's function for oscillator - theory}.

Our aim here is to measure $G(\omega)$ experimentally. This is complicated by the fact that we do not have direct access to $p_B(t)$, the pressure contribution from the bottle, since the signal measured by our microphone will sum the contributions from the speaker (the ``background") and the bottle (the ``signal"):
\begin{equation}\label{eq:general sum}
\underbrace{p_M(t)}_{\text{microphone}} = \underbrace{p_S(t)}_{\text{speaker}} + \underbrace{p_B(t)}_{\text{bottle}}.
\end{equation}
To extract $G(\omega)$, we will therefore need to develop strategies for inferring $p_B(t)$.

\section{Pure Tones}\label{sec:pure tones}

Wilkinson \textit{et al.} \cite{Wilkinson2016} already discussed how to measure $G(\omega)$ using pure tones, but we will revisit that problem in this section with the goal of establishing the basic physical model of Eq.~\ref{eq:general sum}---i.e., our contention that the signal at the microphone $p_M(t)$ is the sum of the speaker signal $p_S(t)$ and the bottle signal $p_B(t)$. As we will see below, this better matches experimental results than the normalization procedure of Wilkinson \textit{et al}.

For pure tones, Eq.~\ref{eq:general sum} gives
\begin{equation}\label{eq:sum of speaker and bottle contributions}
\underbrace{P_{S} \sin(\omega t)}_{\text{speaker}} + \underbrace{P_{B} \sin(\omega t - \delta_B)}_{\text{bottle}} = \underbrace{P_M \sin(\omega t - \delta_M)}_{\text{microphone}}.
\end{equation}
The quantities $P_S$, $P_B$, and $P_M$, in this expression, are real-valued amplitudes. We will take measurements twice---once without the bottle below the microphone, and once with the bottle---to gather enough information to extract these amplitudes and their relative phases.

First, we play a tone of frequency $f$ (i.e., of angular frequency $\omega = 2 \pi f$) without the bottle, collect data, and fit the time-series data. The microphone measures a signal
\begin{equation}
p_\mathrm{no}(t) = P_{S} \sin(\omega t + \phi_{P,\mathrm{no}}),
\end{equation}
corresponding to the voltage driving the speaker
\begin{equation}
v_\mathrm{no}(t) = V_\mathrm{no} \sin(\omega t + \phi_{V,\mathrm{no}}).
\end{equation}
where the ``no" subscripts refer to the fact that no bottle is below the microphone. Note that both the microphone and the speaker voltage have the same time dependence, set by the frequency of the voltage driving the speaker. 

Next, we perform the same experiment with a bottle under the microphone. The microphone measures a signal
\begin{equation}
p_\mathrm{yes}(t) = P_{M} \sin(\omega t + \phi_{P,\mathrm{yes}}),
\end{equation}
corresponding to the voltage driving the speaker
\begin{equation}
v_\mathrm{yes}(t) = V_\mathrm{yes} \sin(\omega t + \phi_{V,\mathrm{yes}}),
\end{equation}
where the ``yes" subscripts indicate that a bottle is below the microphone. (In principle $v_\mathrm{yes}(t)$ and $v_\mathrm{no}(t)$ could be the same, but here we do not assume triggered measurements, so $\phi_{V,\mathrm{no}}$ and $\phi_{V,\mathrm{yes}}$ may differ.)

From these measurements, one can immediately find the phase shift $\delta_M$ of the microphone signal. If the amplitude of the bottle contribution were zero, we would expect the microphone phase shift $\delta_M$ in Eq.~\ref{eq:sum of speaker and bottle contributions} to be zero as well, so we take the phase shift $\phi_{P,\mathrm{no}} - \phi_{V,\mathrm{no}}$ as a $\delta_M = 0$ shift. ((Signal propagation delay causes $\phi_{P,\mathrm{no}} \neq \phi_{V,\mathrm{no}}$.) A second measurement with the bottle allows the shift to be determined from $\phi_{P,\mathrm{yes}} - \phi_{V,\mathrm{yes}}$. The difference between these two measurements, modulo $2\pi$ and subtracted from $2 \pi$, gives us the phase shift of the microphone signal, relative to the speaker signal:
\begin{equation}\label{eq:delta_M}
\begin{split}
\delta_{M} = 2 \pi-\text{mod}\big(&(\phi_{P,\mathrm{yes}}- \phi_{V,\mathrm{yes}}) \\
&-(\phi_{P,\mathrm{no}} - \phi_{V,\mathrm{no}}), 2\pi\big).
\end{split}
\end{equation}

Since $P_{S}$, $P_{M}$, and $\delta_M$ are all measured, we can recast Eq.~\ref{eq:sum of speaker and bottle contributions} in its complex form
\begin{equation}\label{eq:sum of waves, exponential form}
P_{S} e^{i \omega t} + P_{B} e^{i (\omega t - \delta_B)} = P_{M} e^{i(\omega t - \delta_M)}
\end{equation}
to find $P_{B}$ as
\begin{equation}\label{eq:P_B}
P_{B} = \sqrt{P_S^2 + P_M^2 - 2 P_S P_M \cos(\delta_M)}.
\end{equation}
and $\delta_{B}$ as
\begin{equation}\label{eq:delta_B}
\delta_{B} = \arccos((P_M \cos(\delta_M) - P_{S})/P_B).
\end{equation}

\begin{figure}
\centering
\includegraphics[width=0.5\textwidth]{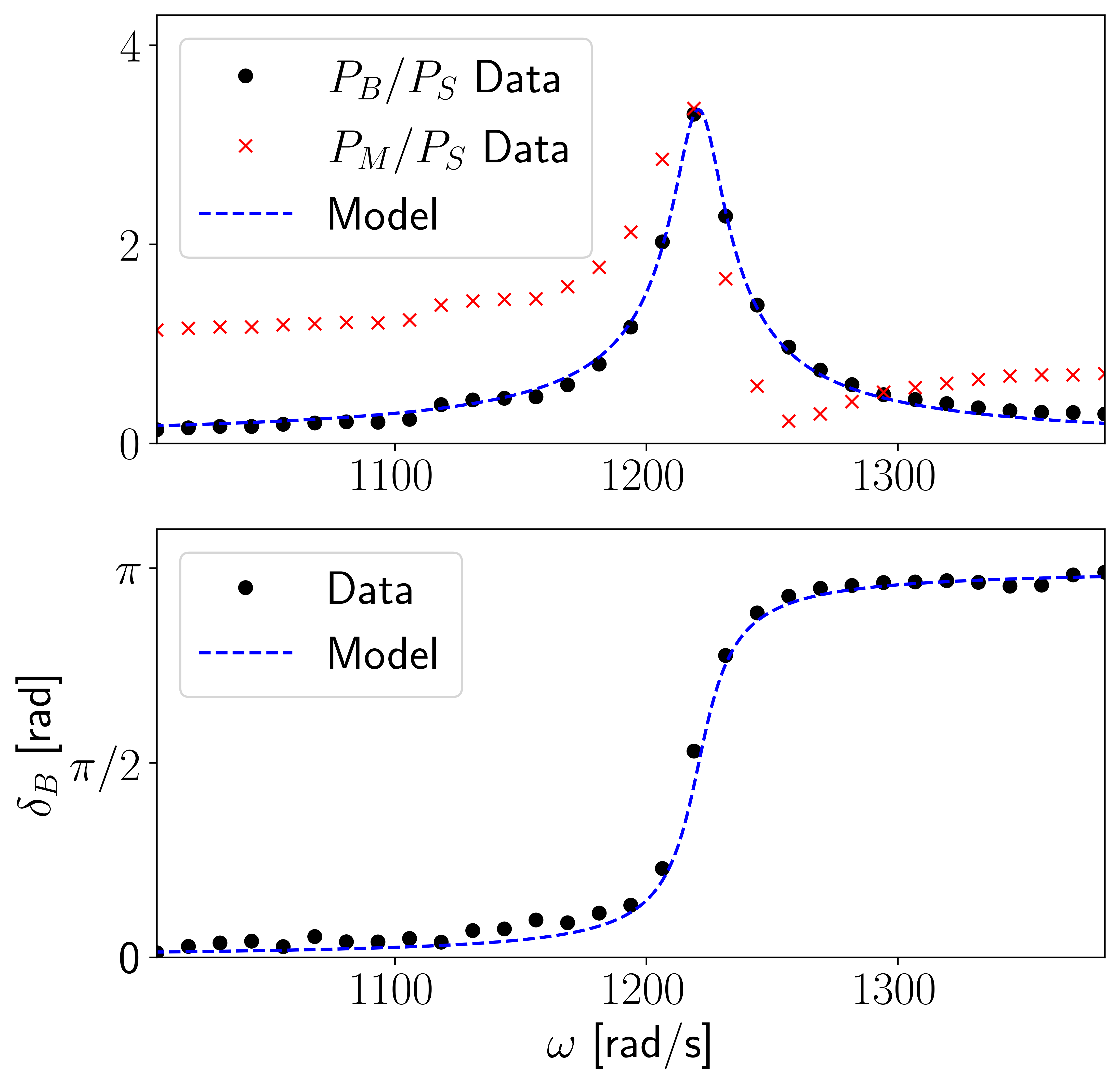}
\caption{Measurements obtained from pure tones at different angular frequencies. \textit{Top}: Normalized amplitude data $P_B/P_S$ and $P_M/P_S$, along with a fit to our model (Eq.~\ref{eq:driven-damped amplitude}). \textit{Bottom}:  Measured phase $\delta_B$ (Eq.~\ref{eq:delta_B}) vs. the fit to our model (Eq.~\ref{eq:G(omega) phase shift prediction}, with the same parameters as for the upper panel).}\label{fig:SHO fit, static data}
\end{figure}

We can perform a nonlinear fit of the measured ratios of $P_B/P_S$ \cite{Ripka2018} to our model prediction 
\begin{equation}\label{eq:driven-damped amplitude}
\frac{P_B}{P_S} = \frac{2 \alpha \beta \omega_0}{\sqrt{(\omega_0^2-\omega^2)^2+4\beta^2 \omega^2}}.
\end{equation}
This fit yields the parameters
\begin{equation}\label{eq:static data parameter fits}
\begin{split}
\alpha &= 3.4 \pm 0.2 \\
\beta &= 10.4 \pm 0.7\, \text{Hz} \\
\omega_0 &= 1220.9 \pm 0.5 \,\text{Hz}
\end{split}
\end{equation}
and the fit vs. data is plotted in Fig.~\ref{fig:SHO fit, static data}. Because the different experiments gave slightly different results for the same bottle, quoted uncertainties have been broadened to reflect this variation. For each parameter, the sample standard deviation for the quoted parameters from each method has been added in quadrature to the formal fitting uncertainty. This procedure leaves the central values unchanged but enlarges the error bars.

Such fits usually require one to include initial guesses for parameters that are nearly correct, so it is worth noting that they can be estimated from the plot alone. $\alpha$ is roughly the maximum value of $P_B/P_S$, $\beta$ is roughly the half-width at half-maximum of the resonance peak, and $\omega_0$ is roughly the frequency where $P_B/P_S$ is a maximum. 

It is also notable, in Fig.~\ref{fig:SHO fit, static data}, that the normalized amplitude $P_M / P_S$, as considered in Wilkinson \textit{et al.}, cannot be fit by the driven-damped oscillator model. There is a clear physical reason for this. While either $P_M / P_S$ or $P_B / P_S$ fits the model fairly well near $\omega_0$, at driving frequencies well below $\omega_0$, the bottle and speaker contributions are almost in phase, leading to constructive interference and larger amplitudes, while at driving frequencies well above $\omega_0$, they are almost out of phase, leading to destructive interference and smaller amplitudes. The mismatch between $P_M/P_S$ and the model is also evident in the tails of Fig.~\ref{fig:SHO fit, static data}, since $P_M/P_S$ tends toward 1 at frequencies far from $\omega_0$, whereas $P_B/P_S$ tends toward 0.

\section{Chirp Tones}\label{sec: chirp tones}

Although the beer bottle's resonance can be characterized using pure tones, collecting enough data to produce Fig.~\ref{fig:SHO fit, static data} is potentially time-consuming. In this section, we discuss how to extract $G(\omega)$ using tones that sweep across the resonance frequency of the bottle. 

As before, it will be useful to normalize our data to account for the non-uniform frequency response of the speaker and microphone. We can use the same setup as in Fig.~\ref{fig:bottle/no bottle setup}, but now trigger the microphone measurements on the time-dependent speaker voltage. 

Up to uncontrolled fluctuations, the input signal should be the same for the two time-dependent measurements---once with the bottle present, and once without it. As before, we will use these measurements to extract $G(\omega)$. We present two ways of doing this. One method, an ``incoherent" approach, only employs the magnitudes of the FFTs; the other, a ``coherent" approach, is also sensitive to phase.

A simple possibility for an audio signal that varies in frequency and time is the ``chirp" function \cite{DigitalSignal}. The linear chirp function starts at a frequency $f_0$ and ends at a frequency $f_1$, interpolating linearly between the two over the duration $T$. The sine function with these properties can be expressed as
\begin{equation}\label{eq:linear chirp}
x(t)=\sin \left[2\pi \left(\frac{c}{2}t^{2}+f_{0}t\right)\right]
\end{equation}
where
\begin{equation}
c=\frac {f_{1}-f_{0}}{T}.
\end{equation}
The benefit of this form is that its Fourier transform $\tilde{x}(\omega)$ has a fairly flat profile in magnitude (though it oscillates in phase). The methods discussed below do not explicitly depend on the form of input signal, as long as the Fourier components close to $\omega_0$ are large enough to avoid uncontrolled fluctuations in the measured $G(\omega)$. 

Fig.~\ref{fig:LoggerPro, sweep} shows the measurements for a linear chirp of uniform amplitude sweeping from 100-300 Hz in 20 s, a signal which can be easily set up using the ``Generate" menu in Audacity. The top red trace represents $p_M(t)$, the signal measured by the microphone when it has a bottle below it. The bottom blue trace represents $p_S(t)$, the signal measured by the microphone without the bottle below it---i.e., the signal arriving from the speaker alone.

\begin{figure}
\centering
\includegraphics[width=0.5\textwidth]{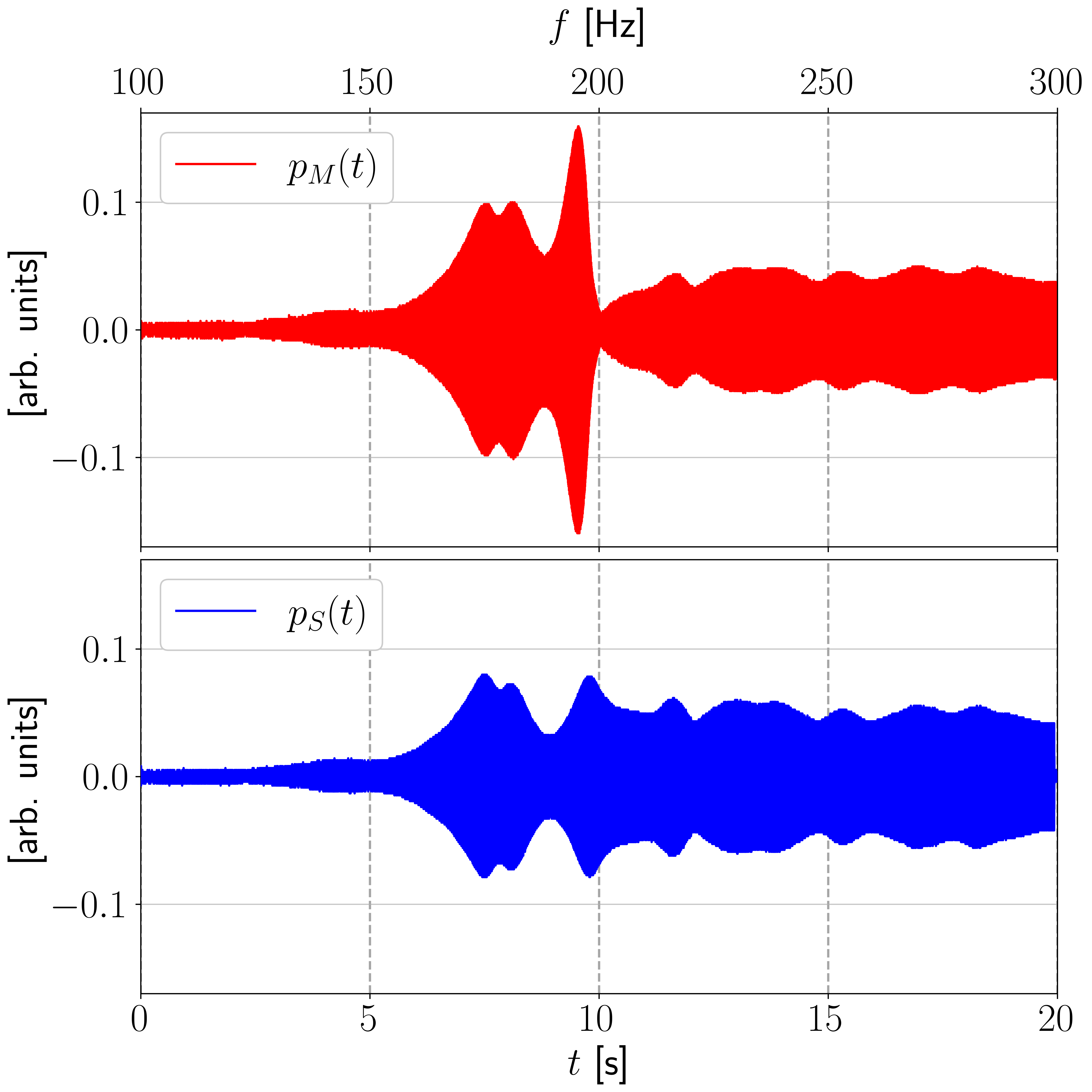}
\caption{Measurements corresponding to a 20.0 s chirp signal sweeping from 100-300 Hz sent by the speaker. \textit{Top}: The microphone signal $p_M(t)$ with the bottle underneath it. \textit{Bottom}: The microphone signal $p_S(t)$ without the bottle below it. For an ideal speaker, the $p_S(t)$ amplitude would be even across all frequencies.}\label{fig:LoggerPro, sweep}
\end{figure}

A few notable things appear in Fig.~\ref{fig:LoggerPro, sweep}. First, while the speaker voltage amplitude is effectively uniform, the shape of the microphone signal without the bottle $p_S(t)$ shows that the speaker's frequency response is markedly inhomogeneous. Nonetheless, $p_M(t)$ shows a distinct amplification around 190 Hz, as we would expect from the previous section, and its variations track those of $p_S(t)$.

\subsection{Incoherent $p_S(t)$ and $p_M(t)$}\label{subsec:incoherent}

To begin our analysis of incoherent signals, we assume the framework introduced in Sec.~\ref{sec:theory}, and take the Fourier transform of Eq.~\ref{eq:general sum}
\begin{equation}
\tilde{p}_M(\omega) = \tilde{p}_S(\omega) + \tilde{p}_B(\omega).
\end{equation}
Using Eq.~\ref{eq:Green's function generic} to write ${p}_B(\omega)$ as $G(\omega) \tilde{p}_S(\omega)$, we find
\begin{equation}\label{eq:pM(omega) in terms of G(omega)}
\tilde{p}_M(\omega) = \big(1 + G(\omega) \big) \tilde{p}_S(\omega).
\end{equation}
We then insert the form of $G(\omega)$ from Eq.~\ref{eq:Green's function specific} to obtain 
\begin{equation}\label{eq:p_M = (1+G in specific form) p_S}
\tilde{p}_M(\omega) = \Bigg( 1  + \frac{2 \alpha \beta \omega_0}{(\omega_0^2 - \omega^2) + 2 i \beta \omega} \Bigg) \tilde{p}_S(
\omega).
\end{equation}

Since we are able to measure $p_M(t)$ and $p_S(t)$---they are the signals, respectively, from the microphone with and without a bottle---we are able to obtain $\tilde{p}_M(\omega)$ and $\tilde{p}_S(\omega)$ via separate FFTs. For this method, we would like to deal only with the magnitudes of these spectra. We can do this by multiplying each side of Eq.~\ref{eq:p_M = (1+G in specific form) p_S} by its own complex conjugate, which gives us
\begin{equation}\label{eq:precursor to R(omega)}
\vert \tilde{p}_M(\omega) \vert^2 = \Bigg(\frac{(2 \alpha \beta \omega_0 + \omega_0^2 - \omega^2)^2 + 4 \beta^2 \omega^2}{(\omega_0^2 - \omega^2)^2 + 4 \beta^2 \omega^2}\Bigg) \vert \tilde{p}_S(\omega) \vert^2
\end{equation}

This suggests a way forward. From the numerical values for $\tilde{p}_M(\omega)$ and $\tilde{p}_S(\omega)$, we calculate the ratio of their squared magnitudes
\begin{equation}\label{eq:R(omega) experiment}
R(\omega) = \frac{\vert \tilde{p}_M(\omega) \vert^2}{\vert \tilde{p}_S(\omega) \vert^2}
\end{equation}
and fit it to the function
\begin{equation}\label{eq:R(omega) theory}
R(\omega) = \frac{(2 \alpha \beta \omega_0 + \omega_0^2 - \omega^2)^2 + 4 \beta^2 \omega^2}{(\omega_0^2 - \omega^2)^2 + 4 \beta^2 \omega^2}.
\end{equation}

A FFT has been applied to each of the signals shown in Fig.~\ref{fig:LoggerPro, sweep}. $\tilde{p}_M(\omega)$ and $\tilde{p}_S(\omega)$ are plotted in the region near the resonance frequency in the upper plot of Fig.~\ref{fig:100-300 Hz spectra, incoherent}, and numerical estimates for $R(\omega)$ and its nonlinear fit in the same frequency range are plotted in the lower plot.

To perform a numerical fit, we need to give initial estimates for the parameters. One way to do this is to numerically estimate the values $\omega_{1}$, where $R(\omega)$ is maximum, and $\omega_2$, where $R(\omega)$ is minimum. Since there are three parameters, we will also need one more value, and can use $R_1 = R(\omega_1)$, the value of $R(\omega)$ at its maximum. Analyzing Eq.~\ref{eq:R(omega) theory}, we find that:
\begin{equation}
\begin{split}
\alpha &\approx \sqrt{R_1 - 1} \\
\beta &\approx \frac{\omega_2^2 - \omega_1^2}{2\omega_1\sqrt{R_1-1}} \\
\omega_0 &\approx \omega_1.
\end{split}
\end{equation}

The spectra and $R(\omega)$ ratio generated from the data shown in Fig.~\ref{fig:LoggerPro, sweep} are displayed in Fig.~\ref{fig:100-300 Hz spectra, incoherent}. Comparing, one can see that the spectra amplitudes roughly follow those of the time-domain signals for the same range of sweep frequencies. Likewise, in this as with the steady-state case, at frequencies below the resonance frequency $\omega_0$, the magnitude of $\tilde{p}_M(\omega)$ is greater than $\tilde{p}_S(\omega)$ due to constructive interference, while above the resonance frequency $\omega_0$, the magnitude of $\tilde{p}_M(\omega)$ is smaller than $\tilde{p}_S(\omega)$ due to destructive interference.

A fit for each pair of $p_M(t)$ and $p_S(t)$ data sets yields a set of parameters. We took five $p_M(t)$ and five $p_S(t)$ datasets, extracted oscillator parameters for each pair (i.e., for $5 \times 5 = 25$ pairs), and averaged, yielding parameter estimates of
\begin{equation}\label{eq:parameter estimates, incoherent}
\begin{split}
\alpha &= 3.0 \pm 0.4 \\
\beta &= 11.0 \pm 0.8\,\text{Hz} \\
\omega_0 &= 1220.4 \pm 0.7\,\text{Hz}.
\end{split}
\end{equation}
The errors have been estimated from the sample standard deviation of each list of 25 estimates, broadened slightly as discussed at the end of Sec.~\ref{sec:pure tones}.

\begin{figure}
\centering
\includegraphics[width=0.5\textwidth]{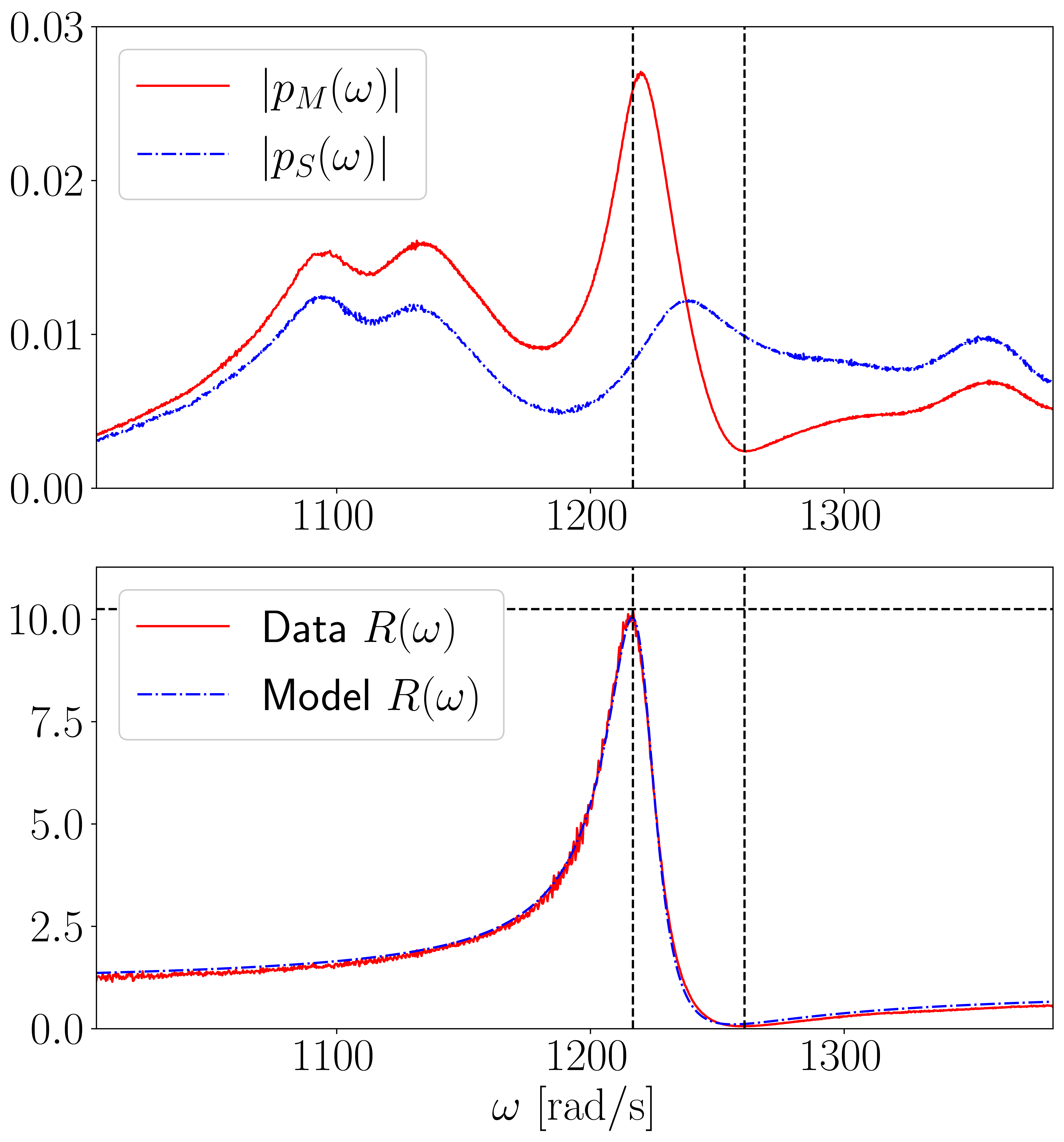}
\caption{\textit{Top}: Spectra of $p_S$ and $p_M$, close to the resonance frequency. \textit{Bottom}: The function $R(\omega)$, with the frequencies $\omega_1$ and $\omega_2$ labeled by vertical dashed lines ($\omega_1 < \omega_2$), and the value $R_1$ labeled by a horizontal dashed line.}\label{fig:100-300 Hz spectra, incoherent}
\end{figure}

Comparing these parameters with those found using the pure tones method (Eq.~\ref{eq:static data parameter fits}), we find that the values are consistent with one another---unsurprisingly, since these measurements were all taken in a single sitting. The parameter $\alpha$ depends sensitively on how far the microphone is from the bottle. The $\omega_0$ and $\beta$ values increase along with increases in temperature, but in our experiments the air varied no more than 1\,\textdegree C between the pure-tones and sweep-tones data acquisition ($T \approx 21.5$\,\textdegree C). 

\subsection{Coherent $p_S(t)$ and $p_M(t)$}\label{subsec:coherent}

In the analysis above, we have assumed that $G(\omega)$ is given by the damped oscillator mode. Here, we develop a method to obtain $G(\omega)$ using only the properties of Fourier transforms \cite{Lemoult2011}. 
 
We will here make use of the convolution theorem. The convolution of two functions, $f(t)$ and $g(t)$, is defined as
\begin{equation}
f*g(t) = \int_{-\infty}^{+\infty} f(t-\tau) g(t) d\tau.
\end{equation}
The convolution theorem establishes that a Fourier transform turns a convolution into a product:
\begin{equation}
\mathscr{F} \left[ f*g(t) \right] = \tilde{f}(\omega) \tilde{g}(\omega).
\end{equation}

We can use this to recover a form for $G(\omega)$. After measuring $p_S(t)$ and $p_M(t)$, we can convolve the time-reversed $p_S(-t)$ with $p_M(t)$, then apply the convolution theorem:
\begin{equation}
\mathscr{F}\left[p_S(-t) * p_M(t) \right] = \mathscr{F}\left[p_S(-t)\right] \mathscr{F}\left[p_M(t) \right]
\end{equation}
The Fourier transform of a time-reversed signal gives us the complex conjugate of the usual transform, so
\begin{equation}
\mathscr{F}\left[p_S(-t) * p_M(t) \right] = \tilde{p}^{*}_{S}(\omega) \tilde{p}_M(\omega).
\end{equation}
If we rewrite $\tilde{p}_M(\omega)$ using Eq.~\ref{eq:pM(omega) in terms of G(omega)}, we find
\begin{equation}
\mathscr{F}\left[p_S(-t) * p_M(t) \right] = \tilde{p}^{*}_{S}(\omega) \tilde{p}_S(\omega) \left(1 + G(\omega)\right).
\end{equation}
Dividing by $\lvert\tilde{p}_S(\omega)\rvert^2$ and subtracting 1 yields
\begin{equation}\label{eq:define F(omega)}
G(\omega) = \frac{\mathscr{F}\left[p_S(-t) * p_M(t) \right]}{\vert\tilde{p}_S(\omega)\vert^2} - 1.
\end{equation}

This is straightforward to calculate numerically. To get $p_S(-t)$, we flip the order of the $p_S(t)$ data, and $p_S(-t) * p_M(t)$ is a numerical convolution of the flipped $p_S(t)$ with $p_M(t)$. The numerator of the fractional term is the result of one FFT, and the denominator is the squared magnitude of another.

\begin{figure}
\centering
\includegraphics[width=0.5\textwidth]{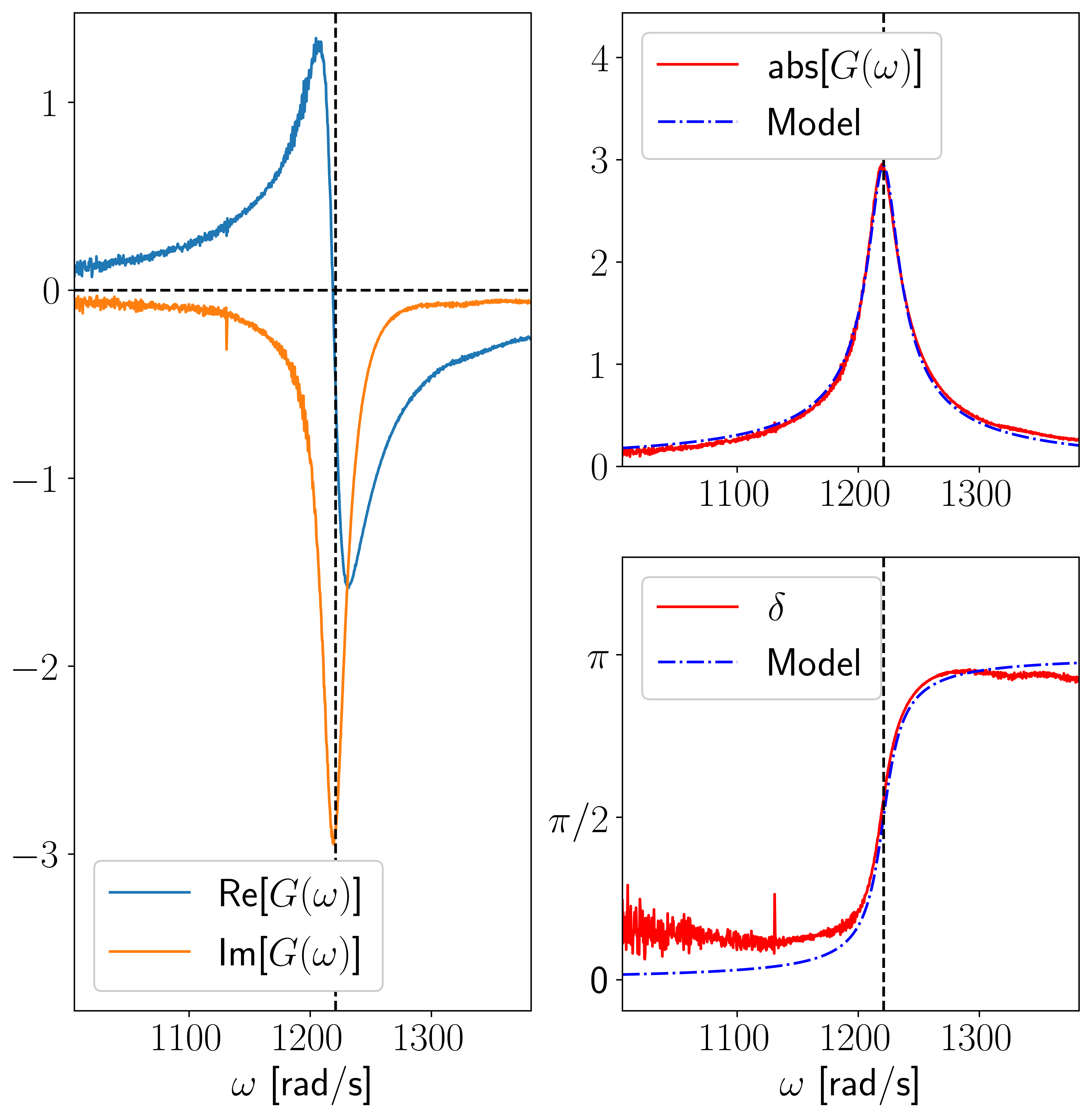}
\caption{$G(\omega)$ from the data shown in Fig.~\ref{fig:LoggerPro, sweep}. \textit{Left}: Real and imaginary parts of $G(\omega)$. \textit{Right top}: Magnitude of $G(\omega)$. \textit{Right bottom}: Phase of $G(\omega)$.}\label{fig:G(omega) reconstruction, sweep}
\end{figure}

This result is shown in Fig.~\ref{fig:G(omega) reconstruction, sweep}, for the same data displayed in Figs.~\ref{fig:LoggerPro, sweep} and~\ref{fig:100-300 Hz spectra, incoherent}. Only a small portion of the calculated estimate for $G(\omega)$ is shown, as outside the frequency range of the sweep tone the Fourier components are not large enough to produce stable quotients. Near the resonance frequency, this gives quite an impressive match with the model expectations set up by Fig.~\ref{fig:green's function for oscillator - theory}, as the model fits for this particular data set make clear.

Using the same five $p_M(t)$ and five $p_S(t)$ datasets as in the incoherent case, and using the same method for estimating parameter errors, we find the parameter estimates
\begin{equation}\label{eq:parmeter estimates, coherent}
\begin{split}
\alpha &= 3.0 \pm 0.4 \\
\beta &= 11.7 \pm 1.2 \,\text{Hz}\\
\omega_0 &= 1221.1 \pm 1.0 \,\text{Hz}.
\end{split}
\end{equation}
Since the same data sets were used for these estimates as in Eq.~\ref{eq:parameter estimates, incoherent}, it is no surprise that they are consistent.

\section{Conclusion}\label{sec:conclusion}

The experiments described above allow a frequency-domain Green's function to be extracted from microphone data. The manipulations are designed to give students experience with FFTs and numerical convolutions at the same time as they explore the physics of driven-damped oscillators. The experiments themselves are very simple to perform, and the beer bottles can easily be switched out for other resonators. It is also possible to use different input signals---e.g., chirps with modulated amplitude, or even more exotic options. More advanced projects might investigate how to describe higher-order harmonics or asymmetric cavities. By allowing students to investigate new resonators with new signals, many other adjacent projects should be possible.

\section*{AUTHOR DECLARATIONS}

\subsection*{Supplementary Materials}

All data files and the Python scripts used to analyze them are available upon request.

\subsection*{Acknowledgments}

This work was supported by the C.M. Hutchinson Family Board of Regents Endowed Professorship for Advanced Scholarship at Centenary College of Louisiana. We thank Kaylin and Caroline Swoboda for donating the beer bottles, and we thank our anonymous peer reviewers for their helpful comments.

\subsection*{Conflict of Interest}

The authors have no conflicts of interest to disclose.

\end{document}